\documentclass[runningheads]{llncs}
\usepackage{algpseudocode}
\usepackage{amsmath,amssymb,amsfonts}
\usepackage{array}
\usepackage{booktabs}       
\usepackage{enumitem}
\usepackage[T1]{fontenc}    
\usepackage{graphicx}
\usepackage{hyperref}       
\usepackage[utf8]{inputenc} 
\usepackage{microtype}      
\usepackage{multirow}
\usepackage[numbers,sort&compress]{natbib}
\usepackage{nicefrac}       
\usepackage{siunitx}
\usepackage{subcaption}
\usepackage{textcomp}
\usepackage{url}            
\usepackage[dvipsnames]{xcolor}

\usepackage{listings}
\usepackage{color}
\usepackage{pifont}

\newcolumntype{L}[1]{>{\raggedright\let\newline\\\arraybackslash\hspace{0pt}}m{#1}}
\newcolumntype{C}[1]{>{\centering\let\newline\\\arraybackslash\hspace{0pt}}m{#1}}
\newcolumntype{R}[1]{>{\raggedleft\let\newline\\\arraybackslash\hspace{0pt}}m{#1}}

\usepackage{tikz}

\makeatletter
\newcommand{\printfnsymbol}[1]{%
  \textsuperscript{\@fnsymbol{#1}}%
}
\makeatother

\begin{document}
\title{AutoPtosis}
%
%
\author{Abdullah Aleem\inst{1,2}\thanks{Equal contribution.} \and
Manoj Prabhakar Nallabothula\inst{1,2}\printfnsymbol{1} \and
Pete Setabutr\inst{1} \and \\ Joelle A. Hallak\inst{1,2} \and Darvin Yi\inst{1,2}}
\authorrunning{Aleem and Nallabothula et al.}
%
\institute{Department of Ophthalmology and Visual Sciences\\ University of Illinois at Chicago \\ Chicago, IL, US
\and Center for AI in Ophthalmology\\ University of Illinois at Chicago \\ Chicago, IL, US\\
\email{\{aaleem2,mnalla2,psetabut,joelle,dyi9\}@uic.edu}}
\maketitle              

\begin{abstract}

Blepharoptosis, or ptosis as it is more commonly referred to, is a condition of the eyelid where the upper eyelid droops. The current diagnosis for ptosis involves cumbersome manual measurements that are time-consuming and prone to human error. In this paper, we present AutoPtosis, an artificial intelligence based system with interpretable results for rapid diagnosis of ptosis. We utilize a diverse dataset collected from the Illinois Ophthalmic Database Atlas (I-ODA) to develop a robust deep learning model for prediction and also develop a clinically inspired model that calculates the marginal reflex distance and iris ratio. AutoPtosis achieved 95.5\% accuracy on physician verified data that had an equal class balance. The proposed algorithm can help in the rapid and timely diagnosis of ptosis, significantly reduce the burden on the healthcare system, and save the patients and clinics valuable resources. 

\end{abstract}

\section{Introduction}

Ptosis (blepharoptosis) \cite{finsterer2003ptosis} is an eyelid condition where the drooping or falling of the upper eyelid causes a narrowing of the palpebral aperture (opening between the eyelids). When the upper eyelid and eyelashes droop, severe vision problems can occur and the quality of life can be affected \cite{federici1999correlation,battu1996improvement,simsek2017association}. In children, severe ptosis can cause amblyopia which results in vision loss if left untreated. Therefore, accurate detection of ptosis followed by appropriate treatment is of vital importance in improving a patient's vision and quality of life.

\begin{figure}[htb]
	\centering
	\includegraphics[width=0.5\textwidth]{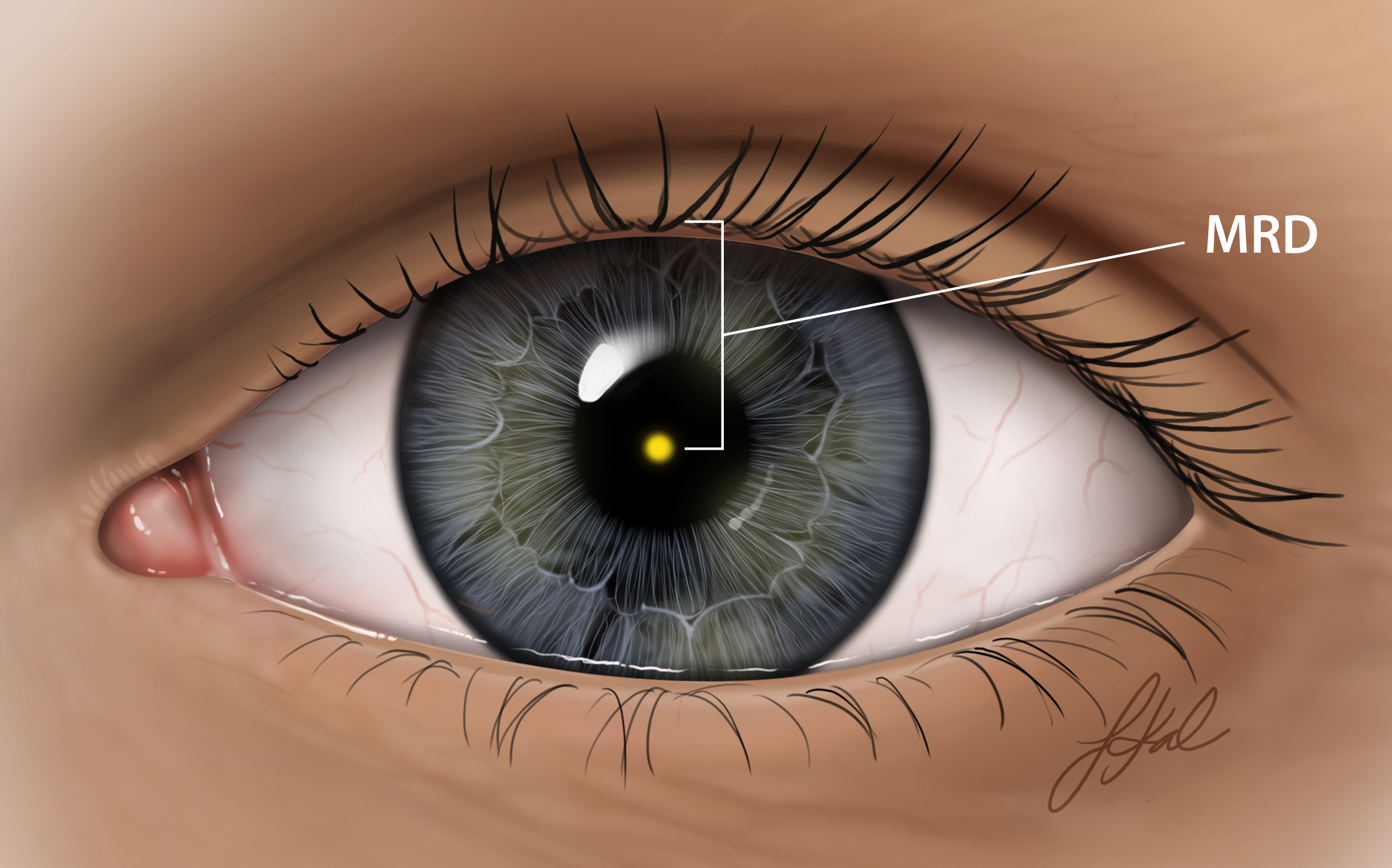}
	\caption{\textbf{MRD1 Distance.}  \small{A depiction of the MRD1 distance measurement.}}
	\label{fig:mrd}
\end{figure}

The current clinical standard for identifying ptosis is by calculating the degree of droop by manually measuring the marginal reflex distance, also known as MRD1 \cite{bodnar2016automated}. MRD1 is the distance between the upper eyelid and the corneal light reflex, as shown in Fig \ref{fig:mrd}. Currently, the methods being using calculate MRD1 rely predominantly on manual measurements. Manual measurements, however, are subjective and prone to human error and are also time-consuming which means the results are not readily accessible for evaluation.

In this paper, we leverage the latest advancements in the field of computer vision and deep learning to develop an automated system for rapid diagnosis of ptosis. This system can be of vital importance for timely detection and treatment of ptosis. It can also help reduce the burden on the healthcare system and save the patients and clinics substantial resources such as time and money.

\section{Related Work}

In recent years, there has been a surge in the usage of teleophthalmology and computer-based applications in the field of ophthalmology as pointed out by Shahbaz and Salducci \cite{shahbaz2020law}. Applications like iScreen, iExaminer \cite{michelson2015learning}, GoCheck Kids Vision Screener \cite{peterseim2018effectiveness} etc., are being widely accepted by the clinicians for their optimal performance in record time and with minimal costs. As mentioned by GoCheckKids, there have not been any computer-based applications that are publicly available for detection of ptosis. The only available solution is iScreen Vision\footnote{\url{https://www.iscreenvision.com/}}, which works only with customized photo screening hardware. 

A recent study by Van Brummen et al. \cite{van2021periorbitai}, also implemented artificial intelligence methods to automate the assessment of periorbital measurements. It used a semantic segmentation network to calculate MRD1 and other distances. In our study, we utilize landmark detection models to calculate the relevant measurement and use them along with a specialized deep learning model which enables automatic disease detection and enhances clinical practices.

\section{Data}

\subsection{Sourcing} The dataset used for this research was sourced from the Illinois Ophthalmic Database Atlas (I-ODA) \cite{mojab2021oda}. I-ODA is a multi-modal longitudinal ophthalmic imaging dataset developed by the Department of Ophthalmology and Visual Sciences at the Illinois Eye and Ear Infirmary of the University of Illinois Chicago (UIC) over the course of 12 years and comprises of patients from different age groups and ethnic backgrounds. The data consisted of 820 full facial images collected in clinical settings for 386 patients that were tested for ptosis. In the sourced data, patients with ptosis outnumbered patients without ptosis because of the increased likelihood of patients with ptosis visiting the hospital for a checkup. Non-ptosis images consisted mostly of postoperative images, however, these images were not available for all patients. Because of the imbalance between ptosis and non-ptosis images we augmented our dataset with 43 facial images of 43 people from the Flickr-Faces-HQ (FFHQ) dataset \cite{karras2019style}, which is a high-quality image dataset of human faces. After augmentation, we had a total of 863 images for 429 unique patients.

\subsection{Labeling}
Patients had ptosis in either a single or both eyes. Hence, four classes (ptosis in both eyes, ptosis in the left eye only, ptosis in the right eye only, and ptosis in no eye) were specified to categorize and label the data. A few images were chosen by an expert physician to represent each category and using these images as a reference, the team manually went over the data and clustered it into the four specified classes. However, we wanted our test data to be as accurate and pristine as possible. For this purpose, we asked an expert physician to hand-select 25 images for each category as precisely as possible. These images created the test set and were used as the ground truth for measuring the performance of our models. The training set was still prone to some errors because the entire dataset was not verified by an expert physician but we approximated labeling to have around 90\% accuracy by comparing the labels of test set from trained physician to what was labeled by our team. The test set consisted of 100 images (77 I-ODA and 23 FFHQ) from 95 people (72 I-ODA and 23 FFHQ) divided into four categories. All images for these 95 people were removed from the original set and what was remaining made up the training and validation set. The training and validation set comprised 656 full facial images (636 I-ODA and 20 FFHQ) from 334 people (314 I-ODA and 20 FFHQ).

\subsection{Pre-processing}
Once the training and testing datasets were finalized, we extracted the region of interest (eye region) from the images using Dlib. Dlib is a face detection and alignment software that is based on the classic Histogram of Oriented Gradients (HOG) feature combined with a linear classifier, an image pyramid, and a sliding window detection scheme developed by Kazemi and Sullivan \cite{kazemi2014one}. Dlib detected 6 landmarks on each eye contour and using these landmarks we extrapolated the area and extracted the eye regions. These regions were then used by our models to predict ptosis. Each extracted eye region was labeled in binary classes, ptosis, and not ptosis, using the four initial categories. After extraction, there were 988 eye images with ptosis and 324 eye images without ptosis.

\section{Methods}

\subsection{Deep Learning}
\begin{figure}[htb]
	\centering
	\includegraphics[width=0.9\textwidth]{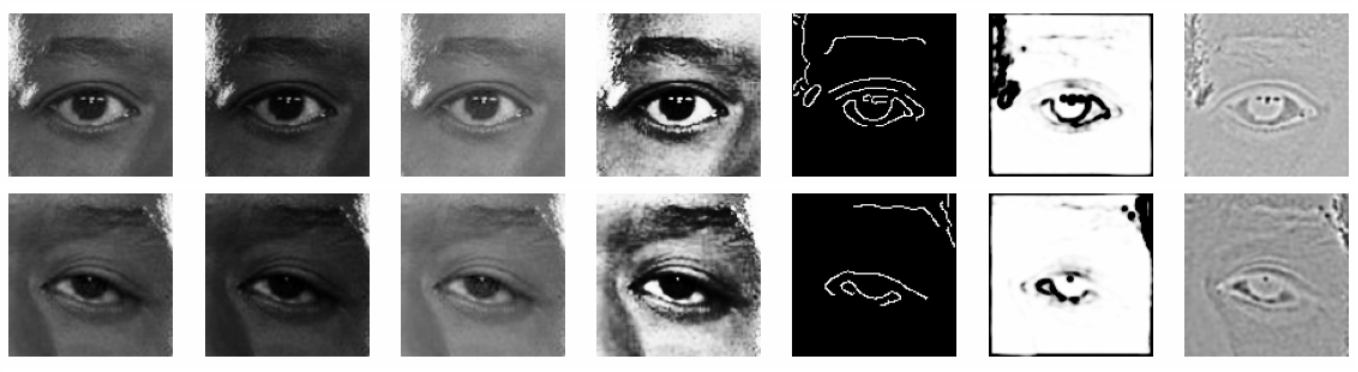}
	\caption{\textbf{Extended Features.}  \small{Eye without ptosis (top) and eye with ptosis (bottom).}}
	\label{fig:extended_features}
\end{figure}

The deep learning model predicted ptosis for each extracted eye region based on the unique characteristics of each class (ptosis and not ptosis) learned during the training phase. Due to limited data available for training, we used a single model for the prediction of both eyes, where the right eye was flipped before being fed into the network. This was based on the assumption that flipping the right eye perfectly translates it to the left eye. To further help the deep learning model we added filtered images alongside the original image. The original image was converted into grayscale and six filtered images were appended as channels. The operations that we used were Gamma correction (with values of 1.5 and 1/1.5), Histogram equalization, Canny edge detection, Harris corner detection, and Difference of Gaussian's. Fig \ref{fig:extended_features} shows an example of the original eye alongside the six appended channels for both, eye without ptosis and eye with ptosis.

Dense Convolutional Network (DenseNet) \cite{huang2017densely}, which connects each layer to every other layer in a feed-forward fashion were used for deep learning. DenseNet121, pre-trained on ImageNet \cite{deng2009imagenet} was fine-tuned using our data. The first and last layers of the model were changed to match our inputs and targets. 100 images (50 from each class) were separated and used for validation and the remaining 1212 images were used for training. The training data had a 3.4 to 1 ratio of ptosis to not ptosis images and a weighted sampler was used to balance the classes. Different hyperparameters were tested and the best performing one (batch size = 32, optimizer = Adam with learning rate of 1e-3 and dropout = 0) were selected using the validation set. The model was trained for at least 100 epochs and stopped after that if there was no improvement in results for 10 epochs and the best model weights were selected based on the validation accuracy. The model was trained 5 times for the given set of hyperparameters and an ensemble of the 5 models was used to minimize the variance of the final model. The output probability values of the 5 models were averaged and a 0.5 cutoff for averaged probability value was used to classify each eye image into ptosis or not ptosis.

\subsection{Clinically Inspired}

The clinically inspired model imitated the clinical procedure and automated the measurements of the MRD1 and Iris Ratio (visible area of the iris) from the eye regions and used those to predict ptosis. The model was divided into three main modules. In the first module, the eye regions that were extracted from Dlib’s predictor were fed into the Mediapipe Iris \cite{ablavatski2020real} model. The Mediapipe Iris model was developed by Google to identify landmarks for the iris, pupil, and eye contours. The Mediapipe Iris model detected the eyelid landmarks (16 points, along the eye contour) and iris landmarks (5 points, along the iris contour) from the extracted eye region. In the second module, the corneal light reflex (CLR) which is the brightest spot in the iris region that is closest to the pupil was extracted from the image. In the last module, the extracted coordinates were used to calculate the MRD1 and IR. The MRD1 was calculated by measuring the distance from detected CLR to the closet coordinate in the upper eye contour and the IR was calculated as a percentage of the iris area inside the eye contour. If the CLR was not visible, the distance from the center of the iris was measured. The MRD1 value was converted from pixels to millimeters based on the assumption that the horizontal iris diameter of the human eye remains roughly constant at 11.7±0.5 mm across a wide population \cite{hashemi2015white, baumeister2004comparison, rufer2005white, bergmanson2017size}. Eventually, to predict ptosis we calculated thresholds for MRD1 and Iris Ratio that maximized the accuracy on our balanced training data and also trained a Decision Tree predictive model. Iris Ratio threshold performed significantly better than MRD1 threshold but Decision Tree outperformed both the individual features and was eventually used in the clinically inspired pipeline for predicting ptosis.

\subsection{AutoPtosis}
Initially, we aimed to combine the deep learning model's probability value and clinically inspired model's MRD1 and Iris Ratio to train a predictive model (logistic regression) for the detection of ptosis. Analyzing the deep learning model's performance on our dataset that contained a wide spectrum of ptosis severity, we noticed that model was rarely wrong when the eye had visibly clear ptosis (severe ptosis) or no ptosis at all. However, for borderline cases where even the physicians had a difficult time classifying them, the model also suffered at times. Hence, using validation data we calculated threshold values of 0.34 and 0.78, below and above which the deep learning model had 100\% accuracy. Using these threshold values we created a combined pipeline where if the deep learning model's probability was below or above these thresholds we used its prediction, otherwise, we used the predictive model.

\begin{figure}[htb]
	\centering
	\includegraphics[width=0.9\textwidth]{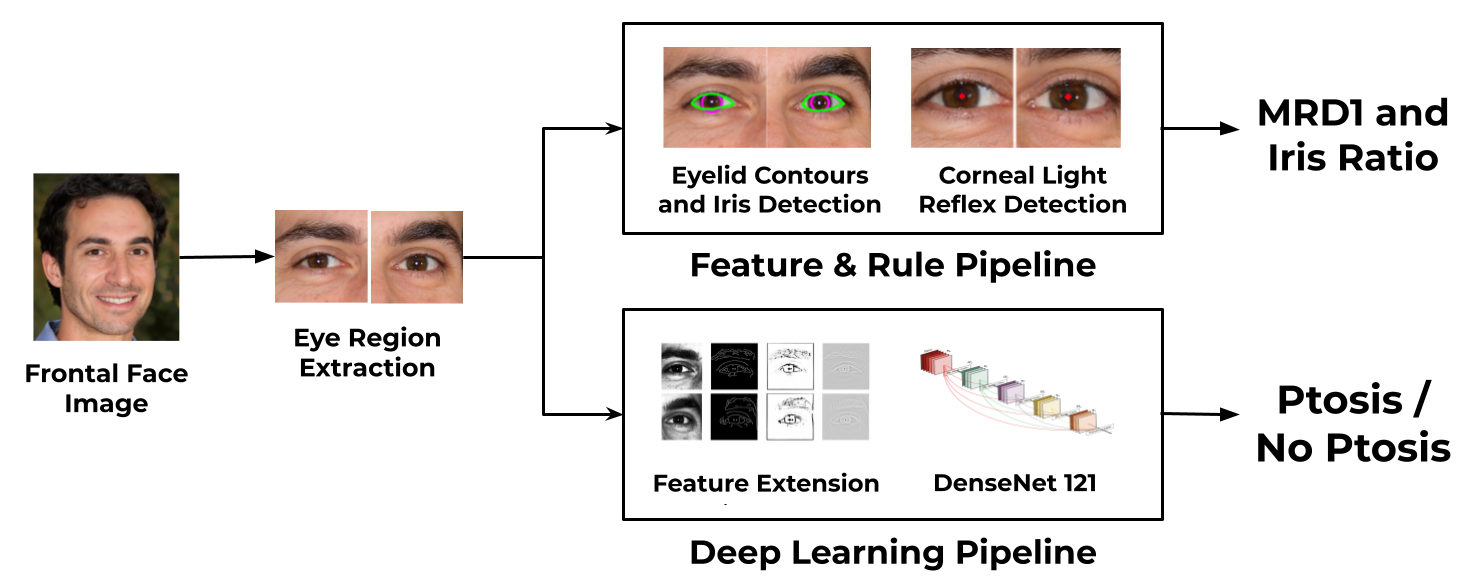}
	\caption{\textbf{AutoPtosis Pipeline.}  \small{A schematic overview of the complete autoptosis pipeline.}}
	\label{fig:autoptosis_pipeline}
\end{figure}

Although, there was a slight improvement in some metric scores after combining both the models the overall accuracy decreased. We reckoned this was due to the substandard performance of the clinically inspired model as compared to the deep learning model. Hence, for AutoPtosis, we decided to use only the deep learning model for the prediction of ptosis. The clinically inspired model's MRD1 and Iris Ratio were output alongside the prediction that helps to determine the severity of the ptosis and further help the clinicians. Fig. \ref{fig:autoptosis_pipeline} shows a complete schematic overview of AutoPtosis.

\section{Results}
\subsection{Quantitative Results}

\begin{figure}[htb]
	\centering
	\includegraphics[width=0.8\textwidth]{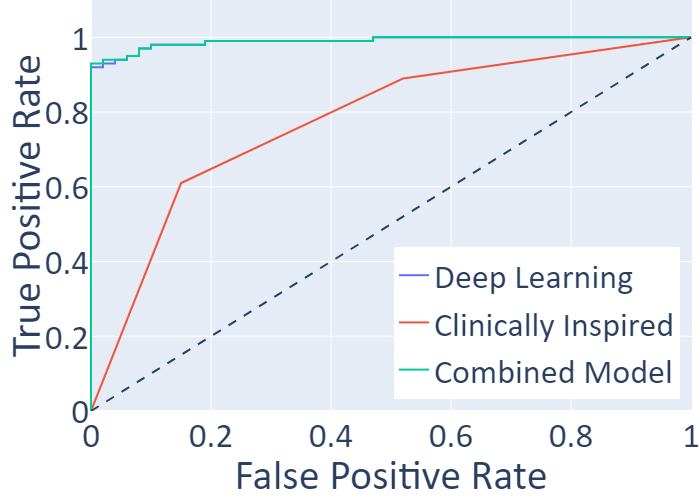}
	\caption{\textbf{ROC Curves Comparison.} \small{}}
	\label{fig:roc}
\end{figure}

Table \ref{table:Comparison} summarizes and compares different performance metrics and Fig. \ref{fig:roc} compares the receiver operating characteristic (ROC) curves for the deep learning, clinically inspired, and combined models.

\begin{table}[htb]
 \caption{\textbf{Performance Comparison.} \small{Percentage scores for different metrics.}}\label{table:Comparison}
\begin{center}
\begin{tabular}{L{2.8cm}|C{01.7cm}C{1.7cm}C{1.7cm}C{1.7cm}C{1.7cm}}
\toprule[1.5pt]
Method&Accuracy&Precision&Recall&F1 Score&ROC AUC\\
\midrule
\midrule
Deep Learning&95.5&97.8&93.0&95.3&98.9\\
Clinically Inspired&73.0&80.2&61.0&69.3&77.6\\
Combined Model&94.0&100&88.0&93.6&99.0\\
\midrule
\midrule
\end{tabular}
\end{center}
\end{table}

The deep learning model outperformed the clinically inspired model, achieving a higher score in all performance metrics. When we combined of both the models using a predictive model and threshold values, we noticed that although the Precision and ROC AUC improved a bit but Accuracy, Recall, and F1 Score deteriorated. The ROC Curves for deep learning and combined models were almost overlapped but the combined model one was faintly better. Overall the deep learning model achieved the best accuracy of 95.5\% on the physician verified test data that had an equal class balance, as compared to the combined model which achieved 94\% and clinically inspired model which achieved 73\%.

Examining the predictions of the clinically inspired model, we noticed that it under-performed for the non-clinical images. We realized that our pipeline often failed to detect the CLR which was either because the eye had drooped to an extent that CLR was not visible anymore or the image was not properly taken. MRD1 was also very sensitive to the detection of upper eyelid contours as even a slight change in pixels resulted in a different prediction. We realized that for this pipeline to work properly, the image needed to be taken in a proper clinical setting and hence removed all FFHQ images from the data and saw an improvement in results as expected. Table \ref{table:FeatureRuleResults} shows the accuracies of MRD1 and Iris Ratio threshold methods as well as Decision Tree for both I-ODA + FFHQ and I-ODA only datasets. There was an improvement in the clinically inspired accuracy for all three methods when only I-ODA images were used. The results for the deep learning model were the same even when FFHQ images were removed which meant that it was robust to such environmental changes and performed well even for images not taken in clinical settings.

\begin{table}[htb]
 \caption{\textbf{Clinically Inspired Accuracy Comparison.} \small{Accuracy comparison for different methods and data sources.}}\label{table:FeatureRuleResults}
\begin{center}
\begin{tabular}{L{4cm}|C{2.5cm}C{2.5cm}C{2.5cm}}
\toprule[1.5pt]
Data Source&MRD1&Iris Ratio&Decision Tree\\
\midrule
\midrule
I-ODA + FFHQ&66.0&71.0&73.0\\
I-ODA&68.0&73.0&79.0\\
\midrule
\midrule
\end{tabular}
\end{center}
\end{table}

\subsection{Qualitative Results}

\begin{figure}[h]
	\centering
	\includegraphics[width=0.8\textwidth]{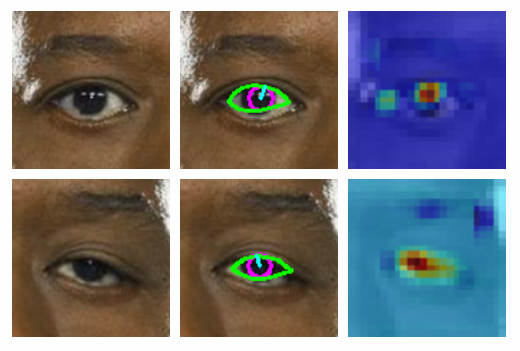}
	\caption{\textbf{Class Activation Map and Direct Feature Visualization} \small{Eye without ptosis (top) and eye with ptosis (bottom).)}}
	\label{fig:cam}
\end{figure}

Interpretability has always been an issue with predictive models. We addressed that problem by developing class activation maps for the deep learning model and direct feature visualization for the clinically inspired model. The class activation maps tell us which regions in the image were most important and contributed the most to the diagnosis. The direct feature visualization shows us the iris and eye contours as well as the corneal light reflex that was used for calculating the MRD1 and Iris ratio. Fig. \ref{fig:cam} shows the right (top) and left (bottom) eye along with the class activation map and direct feature visualization for a person who has ptosis in the right eye only and was diagnosed correctly for both eyes using AutoPtosis. Class activation maps and direct feature visualizations provide a quick and broad visual assessment of the accuracy of the algorithm. We can easily visualize what features of the eye are used to make predictions. In the cases where the heat maps do not correctly show the relevant eye region or the contours are not perfectly drawn on the iris and eyelid, the likelihood of an error is increased and this information can be used by clinics to access the reliability of the models.

\section{Discussion}

AutoPtosis successfully achieved 95.5\% accuracy compared to a trained physician. The automated pipeline significantly reduced the time taken for the diagnosis procedure to a few seconds which was far less than the prior methods. The model was robust and performed well for different age groups and ethnicities owing to underlying diverse dataset. Alongside the prediction, our model also provided the calculated marginal reflex distance and iris ratio which can be used to assess the severity of ptosis and can potentially be used by physicians in determining the appropriate treatment. Class activation maps and direct feature visualizations generated interpretable results which helped in error verification. We hope that these maps and contours would enable a broader acceptance of such automated systems for detection of ptosis in the clinical community. 
\break

Although, the results of AutoPtosis were astounding and achieved in a significantly shorter time frame as compared to traditional methods but like most machine learning algorithms they were not 100\% accurate and completely reliable. Hence, we propose that AutoPtosis should be regarded as a helping tool for clinicians that can aid and improve the existing healthcare system by accelerating the diagnosis process for ptosis and helping in saving valuable resources. We can package AutoPtosis as a web or desktop application which can then be deployed in the clinics where the clinicians can input a properly taken full frontal face image and get a detailed report, which would contain predictions, MRD1, iris ratio, interpretable maps, and statistics e.g. confidence intervals for all values which can then be reviewed by trained physicians. This can help in the prompt and timely diagnosis of disease, eventually resulting in timely treatment and prevent worsening of the condition. It can also significantly reduce the burden on the healthcare system by eliminating the cumbersome manual processes and help clinics save resources such as time and money. AutoPtosis can also be packaged as a mobile application to enable patients in under-served communities without an oculoplastic surgeon to be screened for ptosis. This app can also inform them about potential risk factors, and guide them to the nearby clinics where the reports can be verified by trained physicians and proper treatment can be provided if required. \break

The underlying model of the clinically inspired pipeline, the Mediapipe Iris model, often failed to identify the eyelid and iris contours especially for patients that had ptosis. We assumed that this was because the model we assume was not trained on patients with ptosis. The clinically inspired model also required that the image be taken under proper setup for optimal performance because imaging features such as the presence of corneal light reflex, subjects alignment, and distance of the subject from the camera are critical for correct prediction for MRD1 and iris ratio which are hard to monitor and control in improper setup. Limitations of the underlying model coupled with the failure to often detect CLR resulted in poor prediction of MRD1 and Iris Ratio. Hence, we skipped including these features for the prediction of ptosis as they provided no significant improvement over the deep learning model. However, with improvement in the underlying model which we are planning for future iterations and proper setup, the performance of the clinically inspired pipeline is expected to increase and it might result in improvements of the combined model as well. \break

The detection process of ptosis can also be affected if other conditions are present alongside ptosis that affects the region of interest. Our model failed to correctly detect ptosis for a patient who had arcus senilis (a disease which alters the iris appearance) alongside ptosis. Ptosis can also be confused with other conditions that result in squinting of the eye. Therefore, AutoPtosis is not recommended for anyone who has such alterations and it important to consult a trained physician in these cases.

\section{Future Work}

In the future we aim to create a model specialized for detecting iris and eye contours for ptosis patients using sclera and iris segmentation. This would help us more accurately calculate marginal reflex distance and iris ratio and could also potentially improve the accuracy of the combined model. Also, with availability of more data we aim to include more patients that have other conditions alongside ptosis in our training data to make the model more robust for such cases. More data, will also help in improving the performance of the deep learning model overall and we hope that these changes would further help in assisting the clinicians to accurately predict ptosis.

\section{Conclusion}
AutoPtosis successfully automated the rapid diagnosis of ptosis by leveraging the advancements of computer vision and machine learning. We hope that the 95.5\% accuracy along with the interpretable results would help in broader acceptance of artificial intelligence based systems in healthcare. Advancements in computer vision and machine learning can help us automate and achieve optimal performances in other ophthalmic and medical tasks too. Utilizing such artificial intelligence based systems would enable us to rapidly and timely diagnose diseases, reduces the burden on the healthcare system and can help save the patients and clinics valuable resources.

\clearpage
\bibliography{references}
\bibliographystyle{splncs04}

\end{document}